\documentstyle[12pt,epsfig]{article}
\RequirePackage{tabularx,float,afterpage,pifont}
\RequirePackage{rotating}

\begin{document}

\title{\Large {\bf The Micro Wire Detector}} 

\vspace{1cm}

\hspace{8cm}USC-FP/99-01


\hspace{8cm} January 18, 1998

\vspace{1cm}

\begin{center}

{\Large {\bf The Micro Wire Detector}}

\vspace{.5cm}

\renewcommand{\thefootnote}{\fnsymbol{footnote}}

{\small B. Adeva, F. G\'omez, A. Pazos,\\
R. Pfau\footnote{Visiting student from Heidelberg University.}
, M. Pl\'o\footnote{
Corresponding author: Tel. +34-981-563100-13987, Fax: +34-981-521091,
e-mail: Maximo.Plo@cern.ch}, J. M. Rodr\'{\i}guez, P. V\'azquez}

\vspace{.2cm}

{\small {\it {\bf Universidade de Santiago de Compostela~\footnote{Work supported by Comisi\'on Interministerial de Ciencia y Tecnolog\'{\i}a (CICYT), projects AEN96-1671 and AEN97-1706-E.}}}}

\vspace{.5cm}

{\small J. C. Labb\'e}

\vspace{.2cm}

{\small {\it {\bf CERN}}}

\end{center}

\vspace{1cm}

\begin{center}
{\large Abstract}\\
\vspace{.5cm}
\end{center}

We present the performance of a new proportional gas detector. 
Its geometry consists of a cathode plane with 70x70 $\mu$m$^2$
apertures, crossed by 25$\mu$m anode strips to which it is
attached by 50$\mu$m kapton spacers. In the region where
the avalanche takes place, the anode strips are suspended in
the gas mixture like in a standard wire chamber. 
This detector exhibits high rate capability and large gains,
introducing very little material.

\begin{center}
PACS: 2940, 2970\\
Keywords: MSGC, Gas Detector\\

\vspace{.5cm}
Submitted to Nuclear Instruments and Methods A\\
\end{center}

\newpage

\setcounter{footnote}{0}

\section{Introduction}

A variety of micropattern gas detectors have emerged recently,
in close relation with the introduction of advanced printed
circuit technology. The possibility of kapton etching has allowed 
new geometries like the Gas Electron Multiplier (GEM) \cite{FS1},
the Micro Groove Detector \cite{RB1}, the Well Detector \cite{RB2},
or the Micro Slit Gas Detector (MSGD) \cite{MSD}. These 
avalanche detectors have provided significant improvements
towards the construction of a suitable gas detector for the high
radiation environment of the LHC. We present here, as a result of 
a collaboration between the University of Santiago and the CERN
CMT and SMT groups, a new idea that arises as an improvement
with respect to the MSGD, which has good charge collection
properties, but a limited gain (typically around 2000). 
The openings in the kapton foil have been reduced so as to
have a pattern very similar to a GEM on one side of the
detector layer. The other side, however, is made of metal strips,
or wires, running across the kapton holes.
The better mechanical stability of this setup, having less 
suspended length that in the case of the MSGD, allowed to
produce thinner strips. In this way we obtain a single--stage,
high gain, proportional device, combining the focused electric
field in the micro--hole with the standard wire amplification
and charge collection.
This is why we call this device a Micro Wire Detector ($\mu$WD).
The test of the first prototypes presented here have shown very promising
results of this scheme, as a real option for a high rate tracking 
detector with very low amount of material.

\section{Detector description}

Two prototypes 10 x 10cm$^2$ of the $\mu$WD have been built and tested. 
The first of them consists of a kapton foil 
with a thickness of 50$\mu$m 
copper metallized (5 $\mu$m) on both 
sides. On one side a pattern of square 
holes 70 x 70$\mu$m$^2$
  has been litographically etched. 
On the opposite side 25$\mu$m wide strips  are 
also etched ensuring that they run in the middle of the square holes pattern.
The kapton is then removed in such a way that just an
insulating mechanical joint between anode (strips) and cathode (mesh of
square holes) remains (see Figure 1). The real setup can be appreciated
in the electron microscope photographies shown in Figure 2.
The second detector has been built in the same way, with 60 x 60$\mu$m$^2$
cathode apertures and 25$\mu$m kapton thickness.
The pitch of the anode strips is 100$\mu$m. In this design, the strips are
joined in groups of two at the detector end.
The chamber is finally assembled by enclosing the detector foil between
two 3mm height Vectronite frames, sealed with two kapton metallized foils.
The foil in front of the cathode provides the drift field while the
other is set to ground.

The main differences respect to other micropattern gas detectors
are that anodes are suspended and no substrate for them is needed,
and that these anodes run aligned respect to the 
holes in the cathode metallic mesh.
The detector foil material represents only 0.037\% of a radiation lenght\footnote{In the
complete detector, the contribution on the drift electrode
and gas should be added.}. Moreover
this design allows the construction of a mirror cathode device (see Figure 3)
, using a second kapton foil with another cathode mesh.
This scheme would imply a faster operation by improving the charge
collection time as a consequence of the reduced drift gap. Also,
this configuration would be less sensitive with respect to the 
Lorentz angle under magnetic field.

\section{Detector performance}

The first prototypes have been tested in mixtures of Ar--DME 50--50\%. 
In Figure 4 we present the gain dependence on the cathode and drift
voltages, as measured by a charge sensitive preamplifier\footnote{ORTEC 142PC}
integrating the avalanche charge produced by a 5.9 keV X-ray from a
$^{55}$Fe radioactive source. It can be seen that gains exceeding 
15000 are achievable thanks to the high non--uniformity of the electric
field\footnote{Computed with MAXWELL $3D Parameter Extractor$ program.} in the detector foil as it is shown in Figure 5.

In Figure 6 we show the gain dependence on the cathode voltage
for the two different prototypes. These results were obtained
from X-ray signals from a Cr anode tube. Although the
$\mu$WD with
a kapton thickness of 25$\mu$m exhibits higher gains,
in the present development stage, 
the mechanical and electrical robustness of the 50$\mu$m
foil provided more reliable working conditions.
The operation of the device in non flammable mixtures
(like Ar-CO$_2$ 50--50) is also possible as shown in the
same figure, although with reduced values of the gain factor. 

In order to 
study the rate capability and current distributions,
the detector was
irradiated again with a high intensity Cr X-ray tube.
Rate capability was tested directly using a current amplifier\footnote{
ORTEC VT120} at 
high rate where the peak value from the current spectra 
was monitored. The gain variations are
less than 5\% up to rates as high as 4$\times$10$^5$ Hz mm$^{-2}$
(Figure 7). 

Begining from a cold start, charging up from the kapton spacers
affects the gain less than a 10\%, as shown in Figure 8. The
uniformity of the gain was also measured in 2mm steps over
a lenght of 5cm along the strips.
The variations with respect to the mean were less than 10\% (Figure 9). 
These results  
show that the cathode and anode planarity and the thickness uniformity
of the kapton spacers are good enough. 

 In Figure 10
we show the anode, cathode and drift currents versus the 
cathode voltage (a) and versus the drift field (b) while
the detector was irradiated with a high intensity X-ray beam
with a 2mm diameter collimation. It is possible to 
obtain field configurations in which 90\% of the ions migrate to
the cathode producing a fast charge collection device.
In Figure 11 we show the fast avalanche signal from a current
amplifier (VT 120) originated by a 5.9 keV photon.

\section{Conclusions}

We present a new gas proportional device, the Micro Wire Detector.
This high granularity (100$\mu$m pitch) position sensitive
detector exhibits excellent performance characteristics:
high rate capability (up to .4 10$^6$ Hz/mm$^2$), very low
amount of interposed material (0.037\% X$_0$) and high gain factor ($\sim$10$^4$).
Although further tests and improvements are needed, we 
consider it as a very promising new kind of micropattern
gas device.

\section{Acknowledgements}

One of us (J.C. Labb\'e) would like to thank A. Placci (CERN/EP/TA1)
for his encouragement and assistance to the technical development 
of the detector.

We thank 
A. Monfort, M. S\'anchez (CERN EP/PES/LT Bureau d' etudes Electroniques), L. Mastrostefano and D. Berthet (CERN EST/SM
Section des Techniques Photom\'ecaniques).
Also we acknowledge A. Gandi, responsible of the Section des Techniques
Photom\'ecaniques, for his logistic support.

We are also grateful to Luciano S\'anchez, from LADICIM (Universidad
de Cantabria), for the excellent electron microscope pictures.

\newpage

{\large {\bf Figure Captions}}

\vspace{.5cm}

\begin{tabular}{p{2cm}p{11cm}}

Figure 1: & Design of the Micro Wire Detector foil. \\

Figure 2: &
Electron microscope image of the detector foil
as seen from the cathode side (a), and the same 
foil seen from the anode side (b).\\

Figure 3: & Schematic view of the double cathode
 $\mu$WD  proposed in the article. \\

Figure 4: & Gain of the 50$\mu$m $\mu$WD as a function of the
cathode voltage in Ar-DME 50-50, obtained from 
the pulse height spectra of a $^{55}$Fe source (a).
Gain of the same prototype as a function of drift field (b). \\

Figure 5: & Electric field configuration for one detector cell
in the plane transverse to the anode direction
(across the middle of one hole).
Lines corespond to equal electric field intensity in kV/cm 
(V$_{anode}$=0V, V$_{cathode}$=-500V, V$_{drift}$=-3000V).
Straight lines indicate the limits of the kapton spacer
not intersected by the chosen xz plane.\\

Figure 6: &  Comparison of the callibrated gain factor
between the 50$\mu$m and 25$\mu$m prototypes tested
in Ar-DME 50--50. Also it is shown the gain for the 50$\mu$ prototype
with Ar--CO$_2$ 50--50.\\

Figure 7: & Relative values 
of the
peak current spectra (from the signal of a VT120 
amplifier) versus the photon rate interactions from
a Cr anode X-ray tube.\\

Figure 8: & Relative variations of the gain as a function 
of time, measured every 30s, beginning from a cold start.\\

Figure 9: & Relative variaations of the detector gain 
measured in 2mm steps over a lenght of 5cm along the strips.\\

Figure 10: & Anode, drift and cathode currents versus: 
(a) cathode voltage and (b) drift field,  under high intensity
X-ray irradiation (2mm diameter collimator). \\

Figure 11: & Avalanche signal of a 5.9 keV photon interaction
from a $^{55}$Fe source obtained
with the VT120 ORTEC amplifier. Note that the horizontal scale 
is 10ns/division and vertical scale is 20mV/division.\\

\end{tabular}

\newpage

\pagestyle{empty}

\voffset=-2cm

\begin{figure}
\begin{center}

\vspace{2cm}

\begin{turn}{-90}
\mbox{\epsfig{file=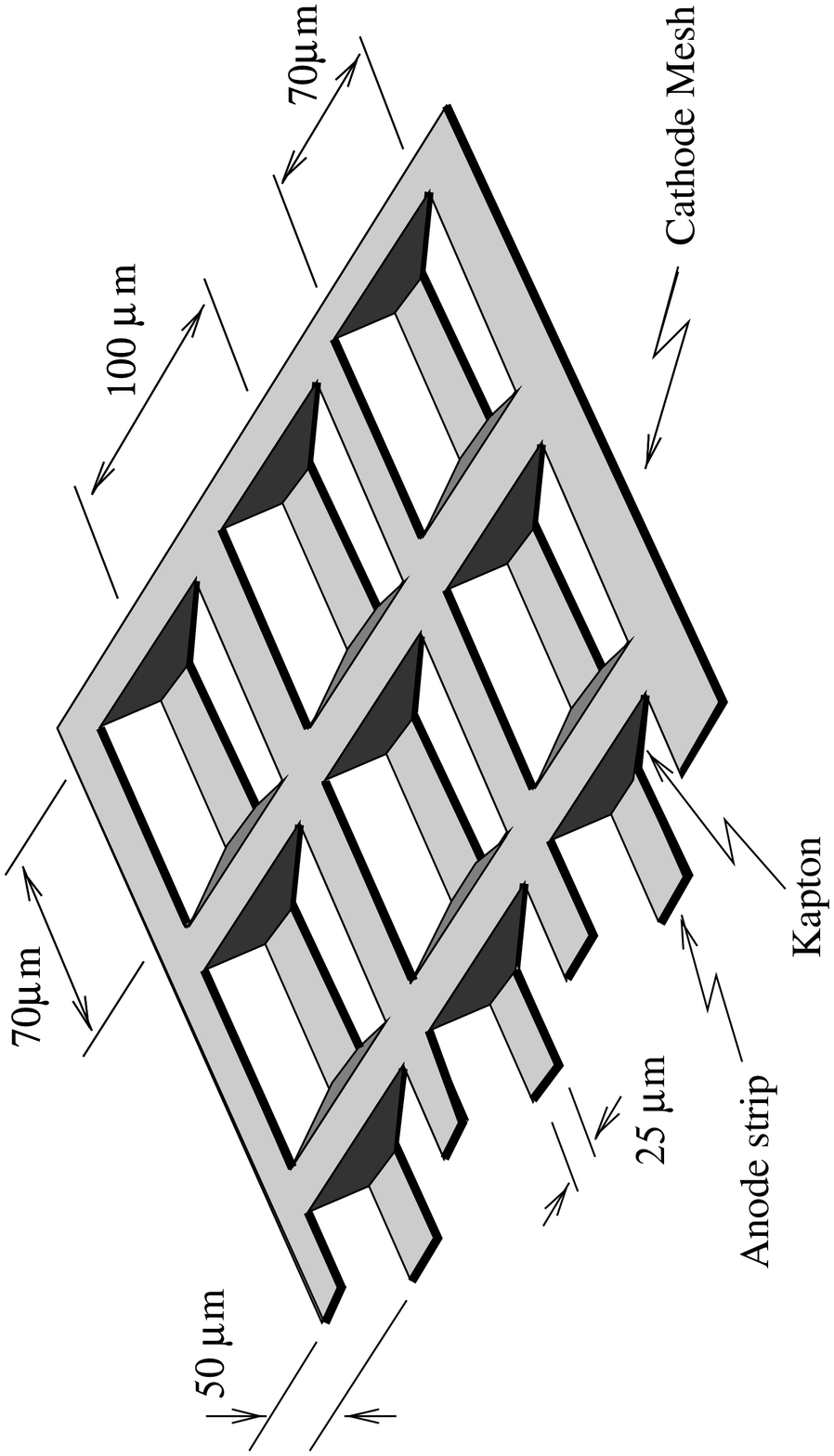,height=12cm}}
\end{turn}

Figure 1

\vspace{1cm}

\begin{tabular}{m{.5cm}m{6cm}}

a) & \mbox{\epsfig{file=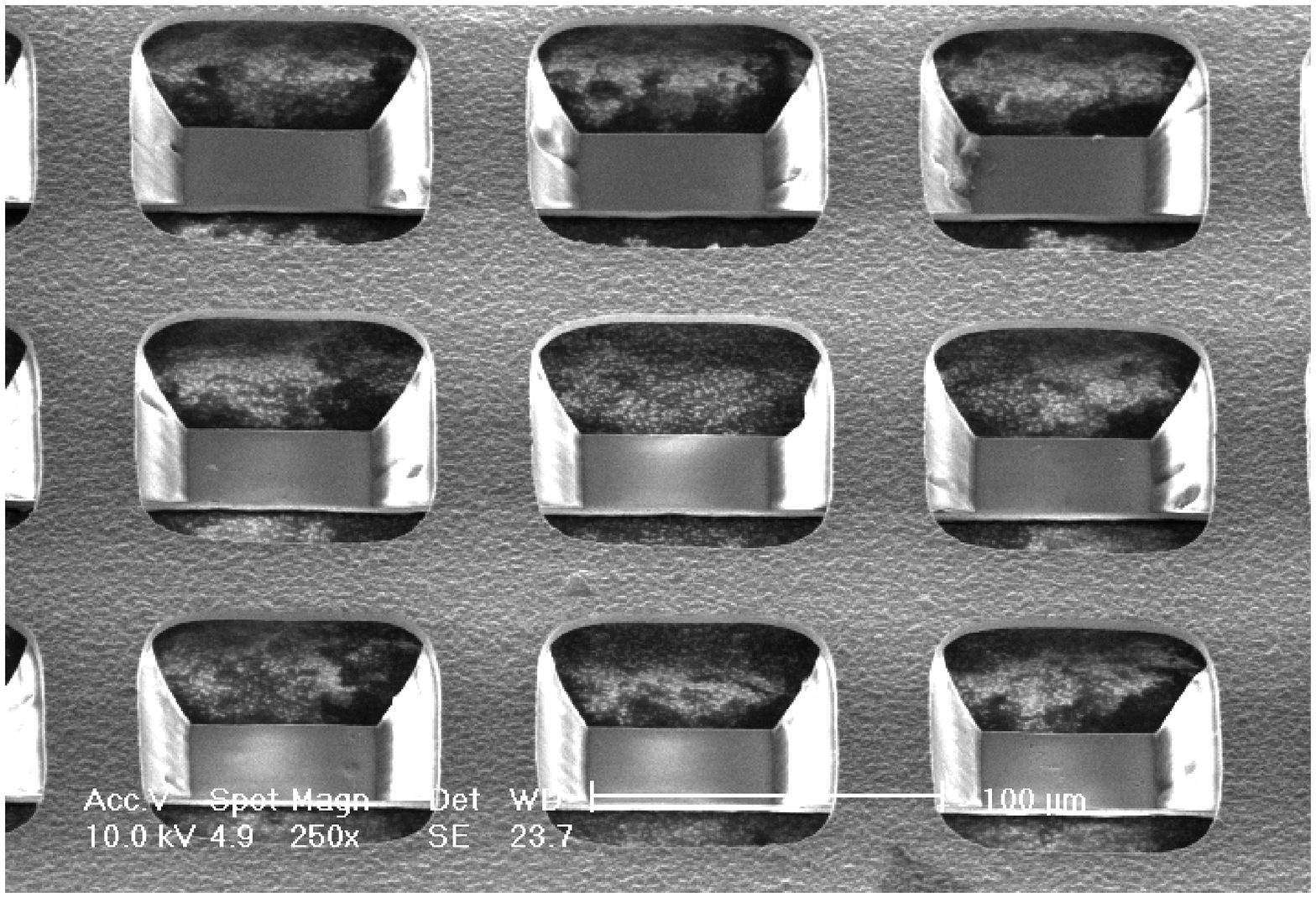,height=6.cm}} \\

b) & \mbox{\epsfig{file=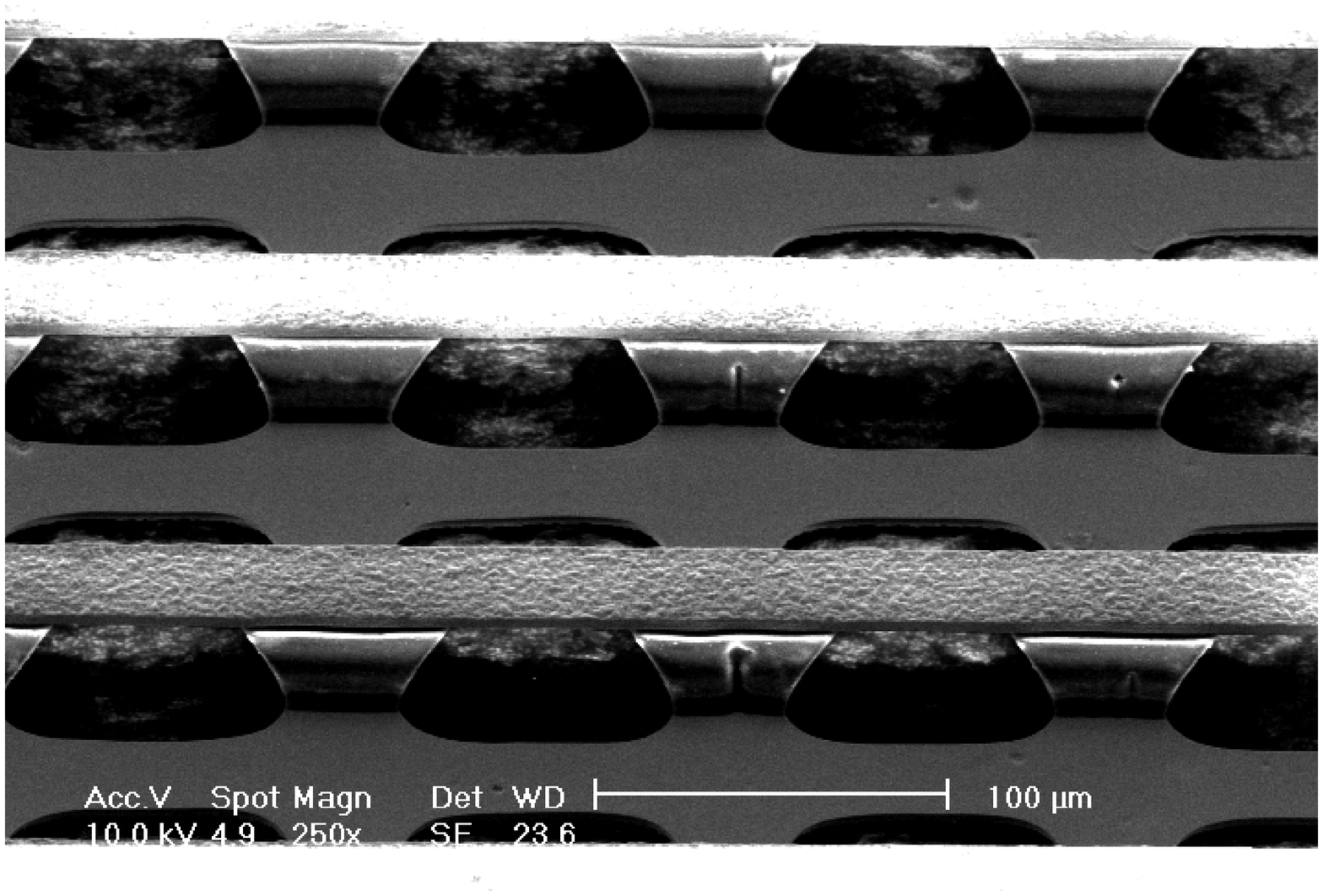,height=6.cm}}
\end{tabular}

\vspace{.5cm}

Figure 2
\end{center}
\end{figure}

\newpage

\vspace{3cm}

\begin{figure}
\begin{center}

\begin{turn}{-90}
\mbox{\epsfig{file=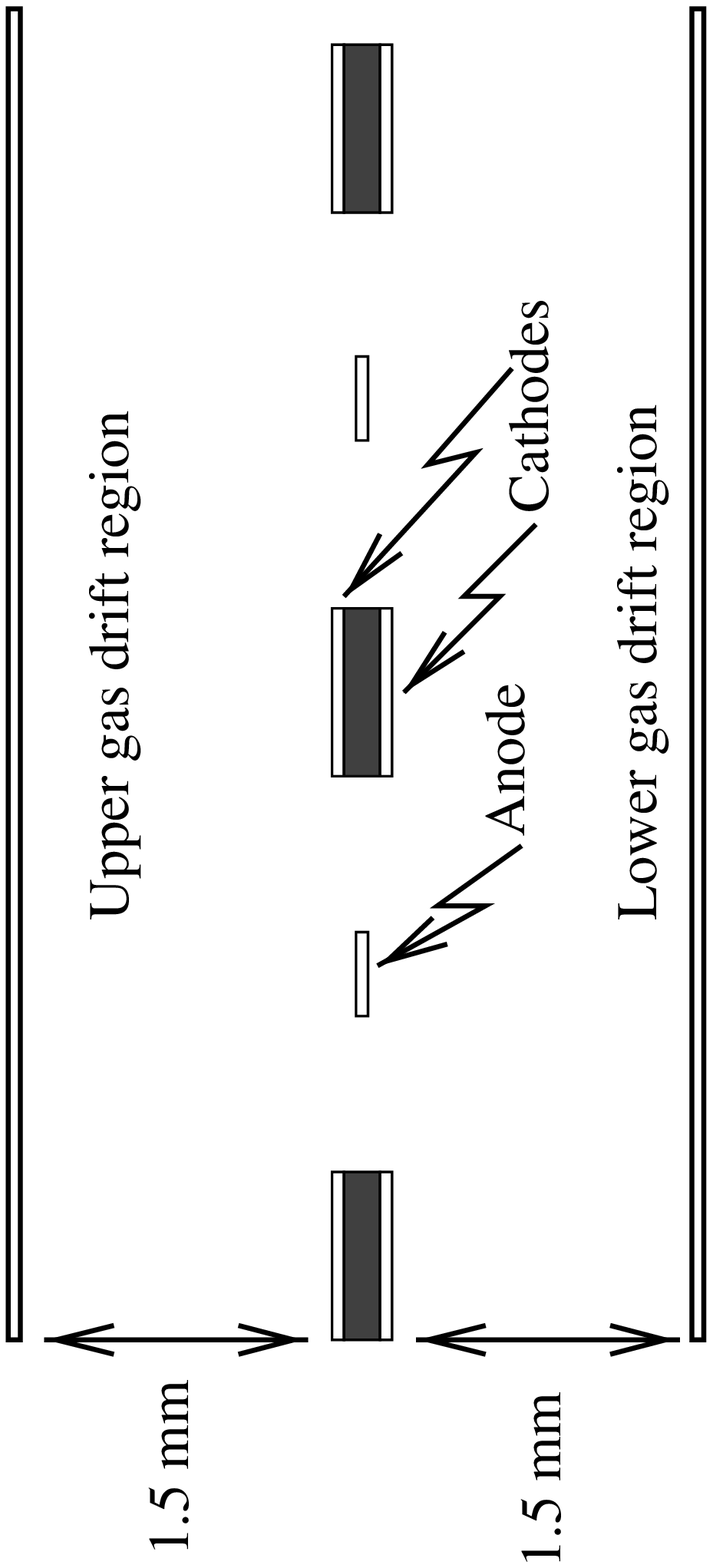,height=10cm}}
\end{turn}

\vspace{.5cm}

Figure 3

\vspace{.5cm}

\begin{tabular}{m{.5cm}m{8cm}}
a) &
\mbox{\epsfig{file=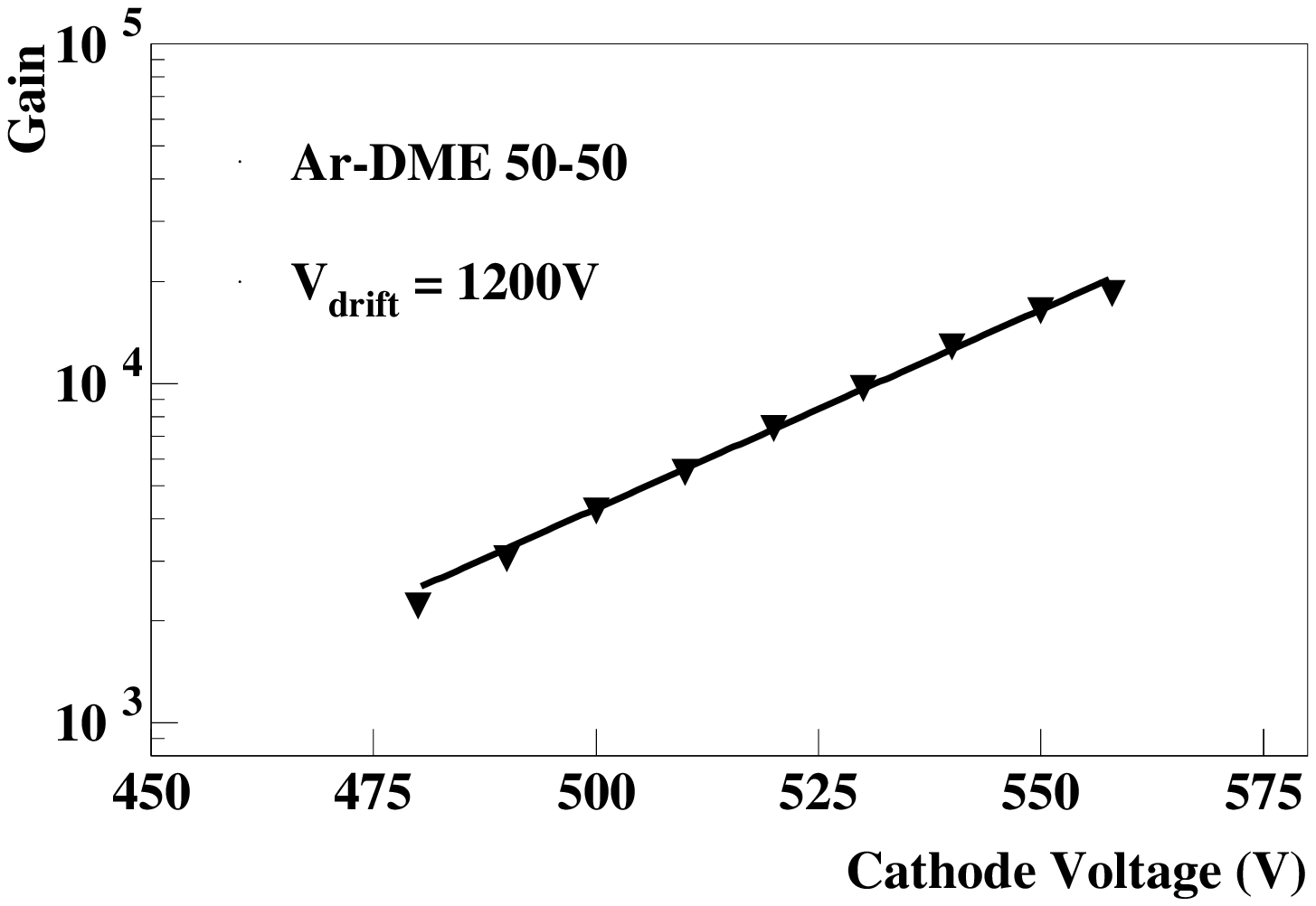,height=8cm}}\\
b) &
\mbox{\epsfig{file=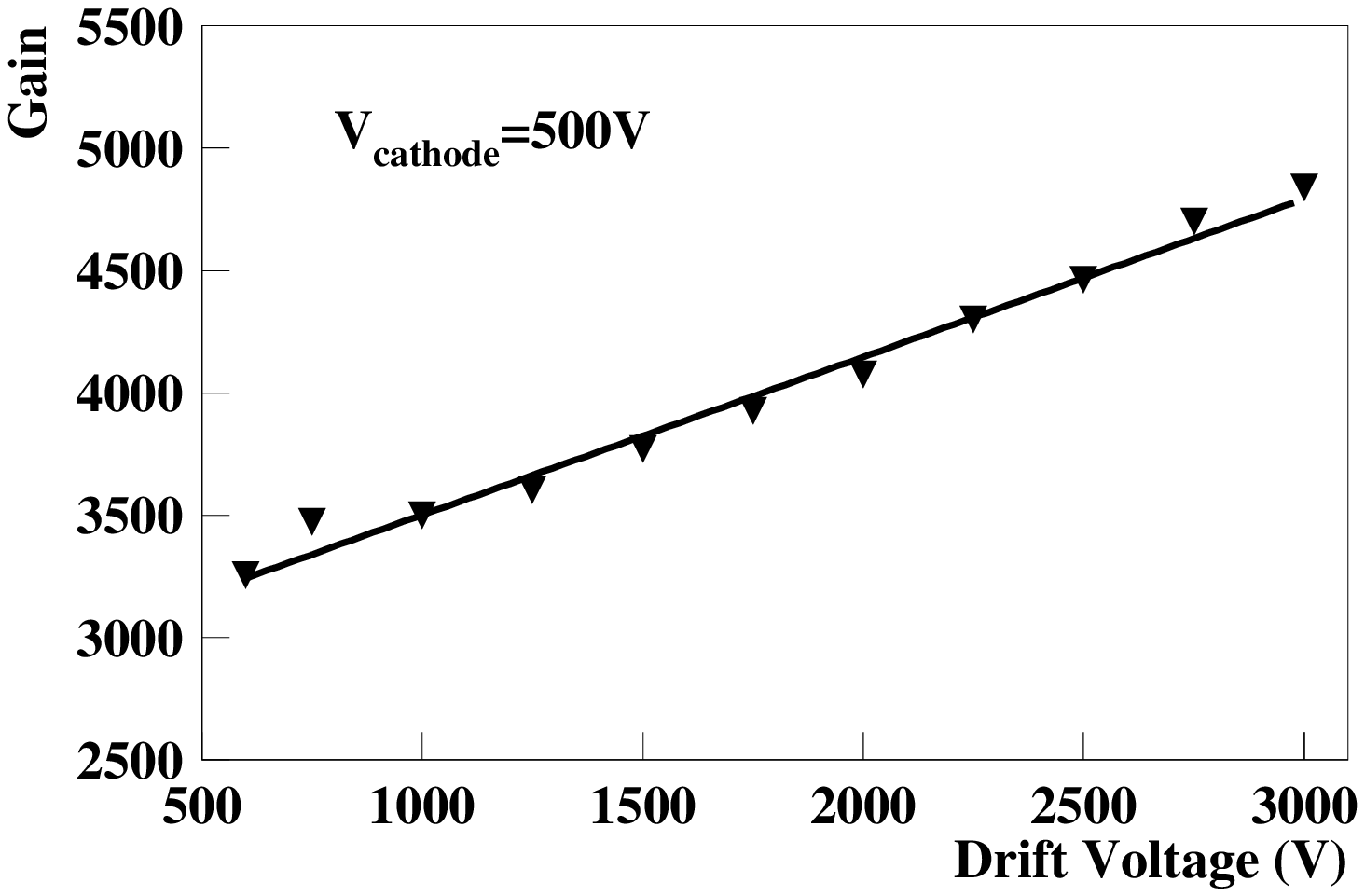,height=8cm}}
\end{tabular}

Figure 4
\end{center}
\end{figure}

\newpage

\begin{figure}
\begin{center}

\begin{turn}{-90}
\mbox{\epsfig{file=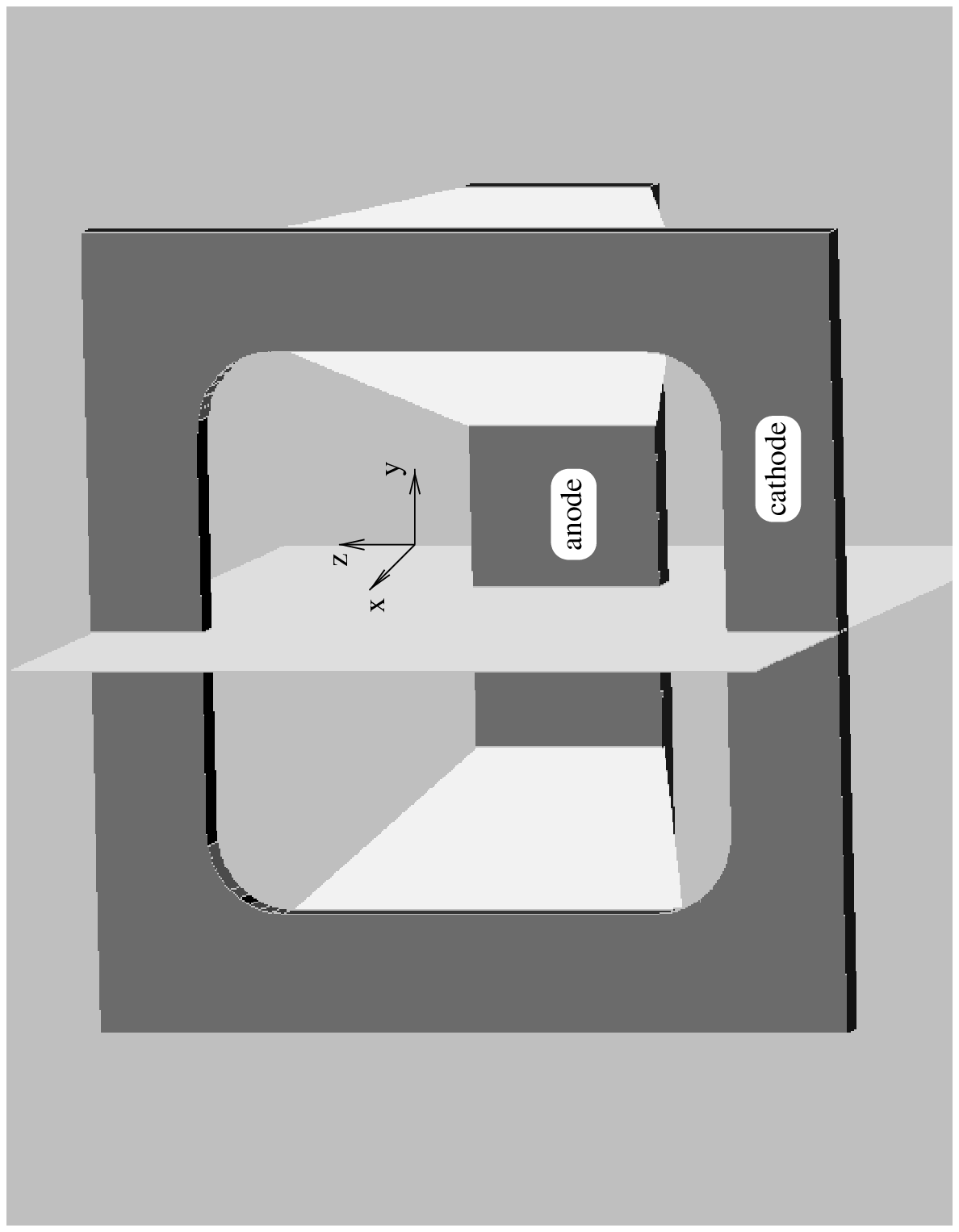,height=10cm}}
\end{turn}

\begin{turn}{-90}
\mbox{\epsfig{file=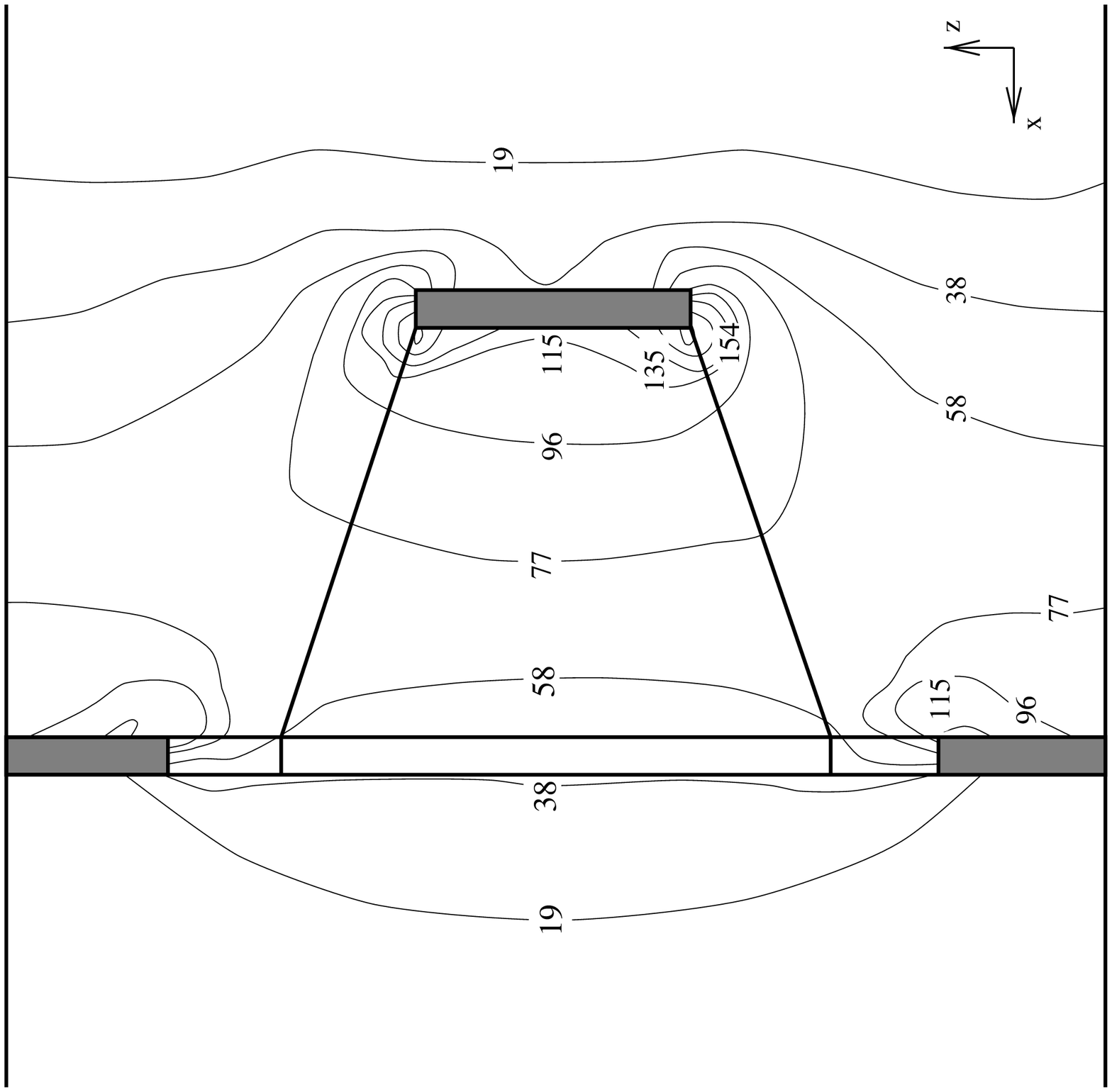,height=10cm}}
\end{turn}

\vspace{1cm}

Figure 5

\end{center}
\end{figure}

\newpage

\begin{figure}
\begin{center}
\mbox{\epsfig{file=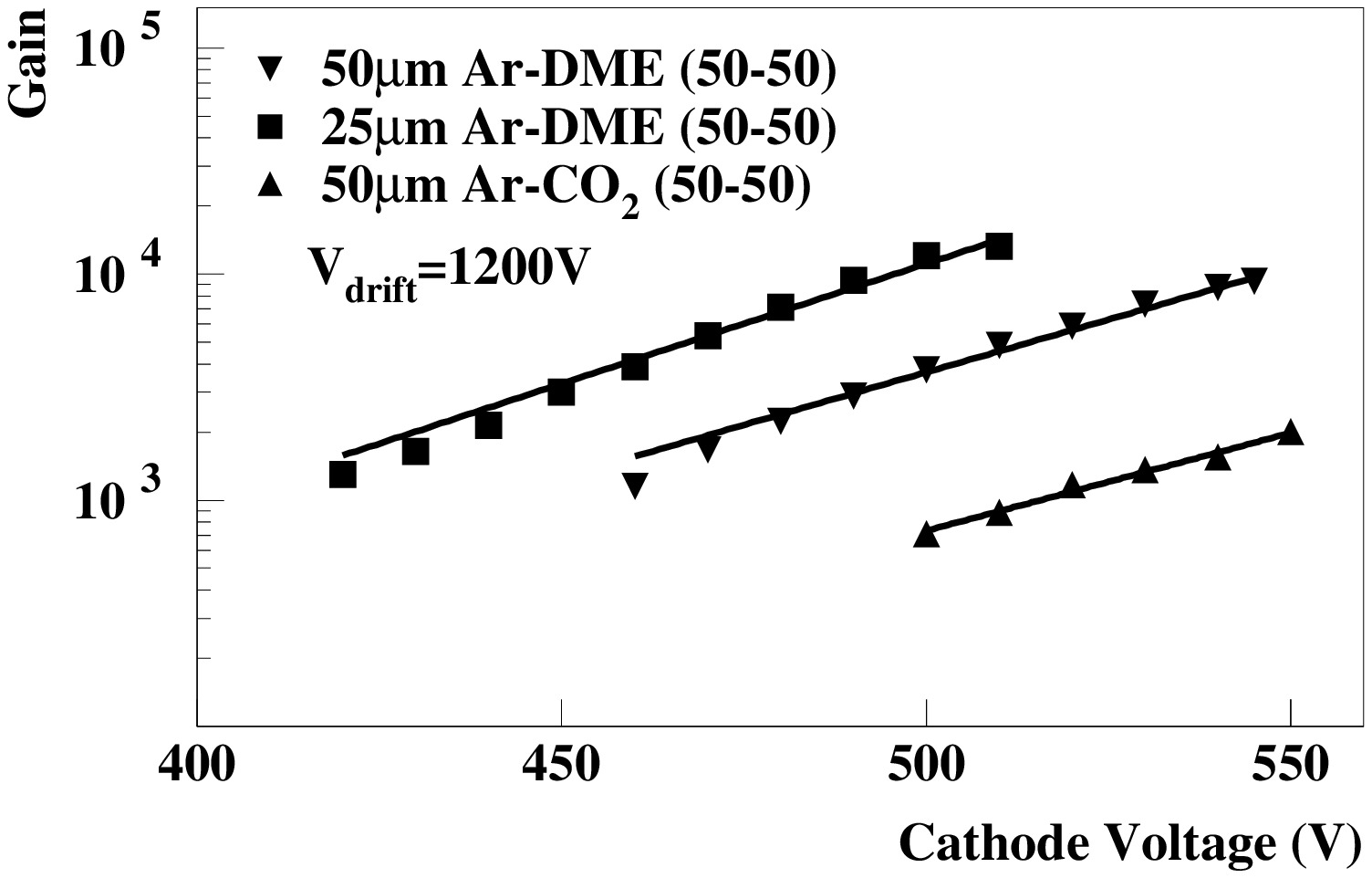,height=8cm}}

Figure 6

\vspace{.5cm}

\mbox{\epsfig{file=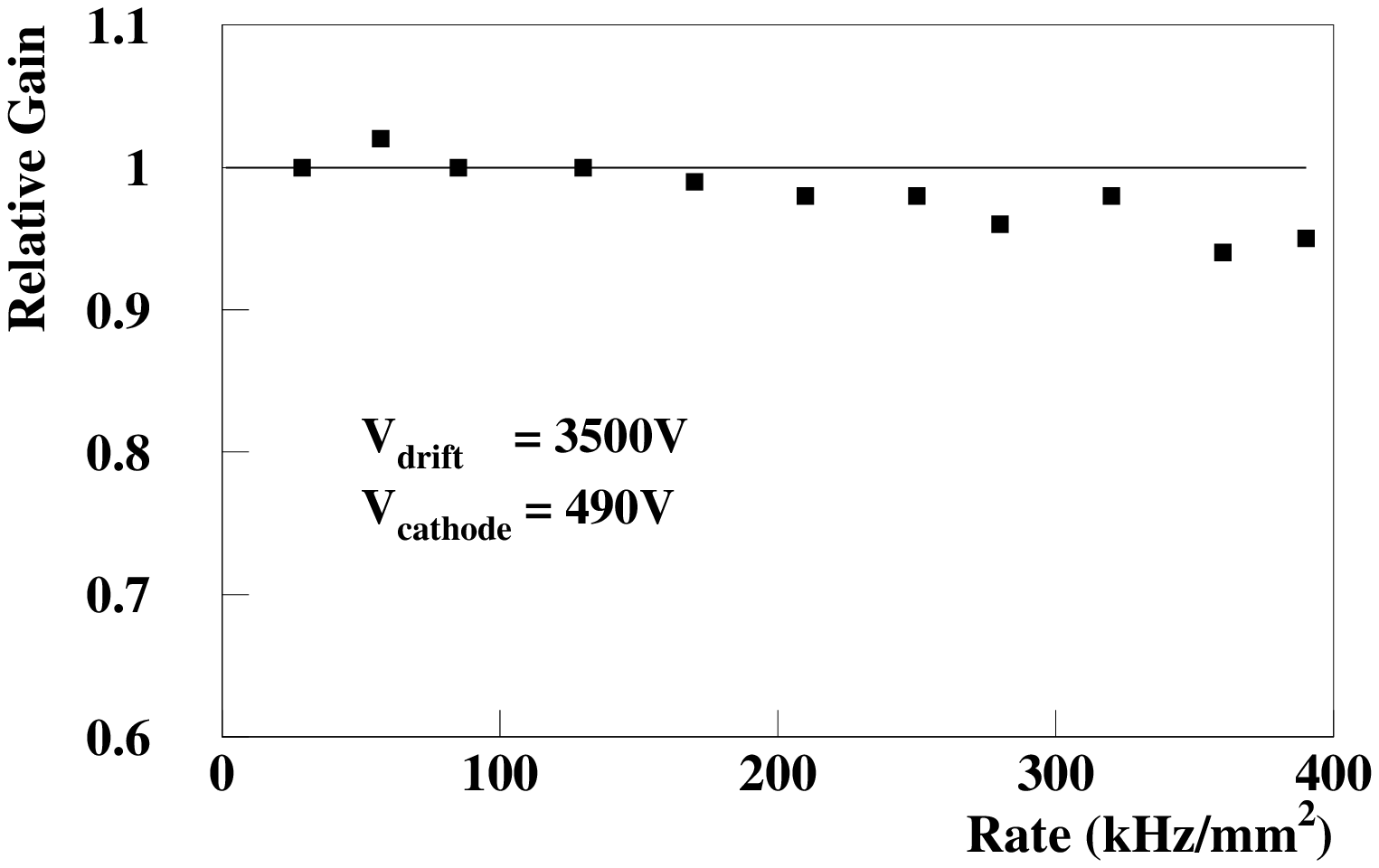,height=8cm}}

Figure 7

\end{center}
\end{figure}

\newpage

\begin{figure}
\begin{center}

\mbox{\epsfig{file=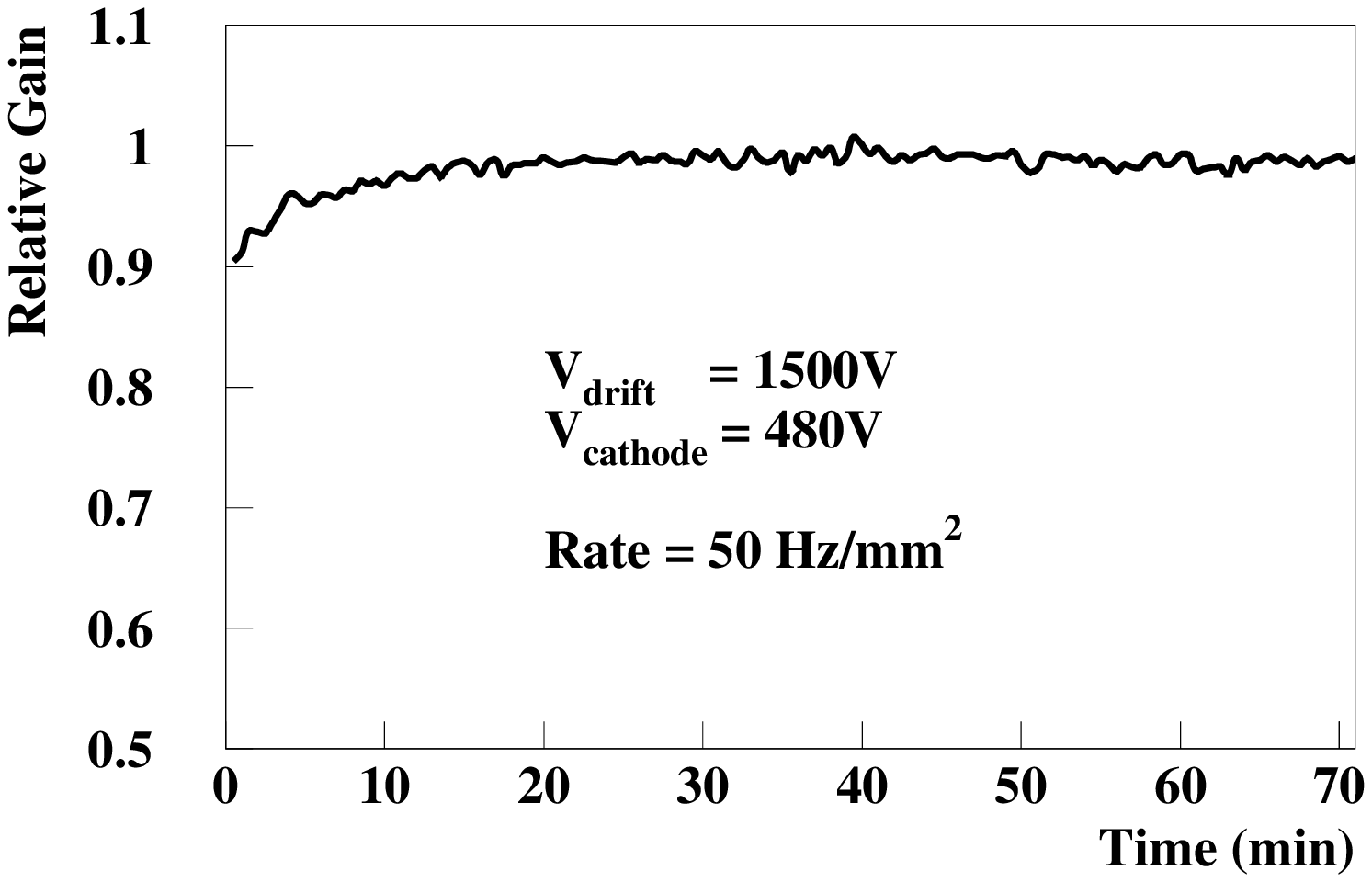,height=8cm}}

Figure 8

\vspace{.5cm}

\mbox{\epsfig{file=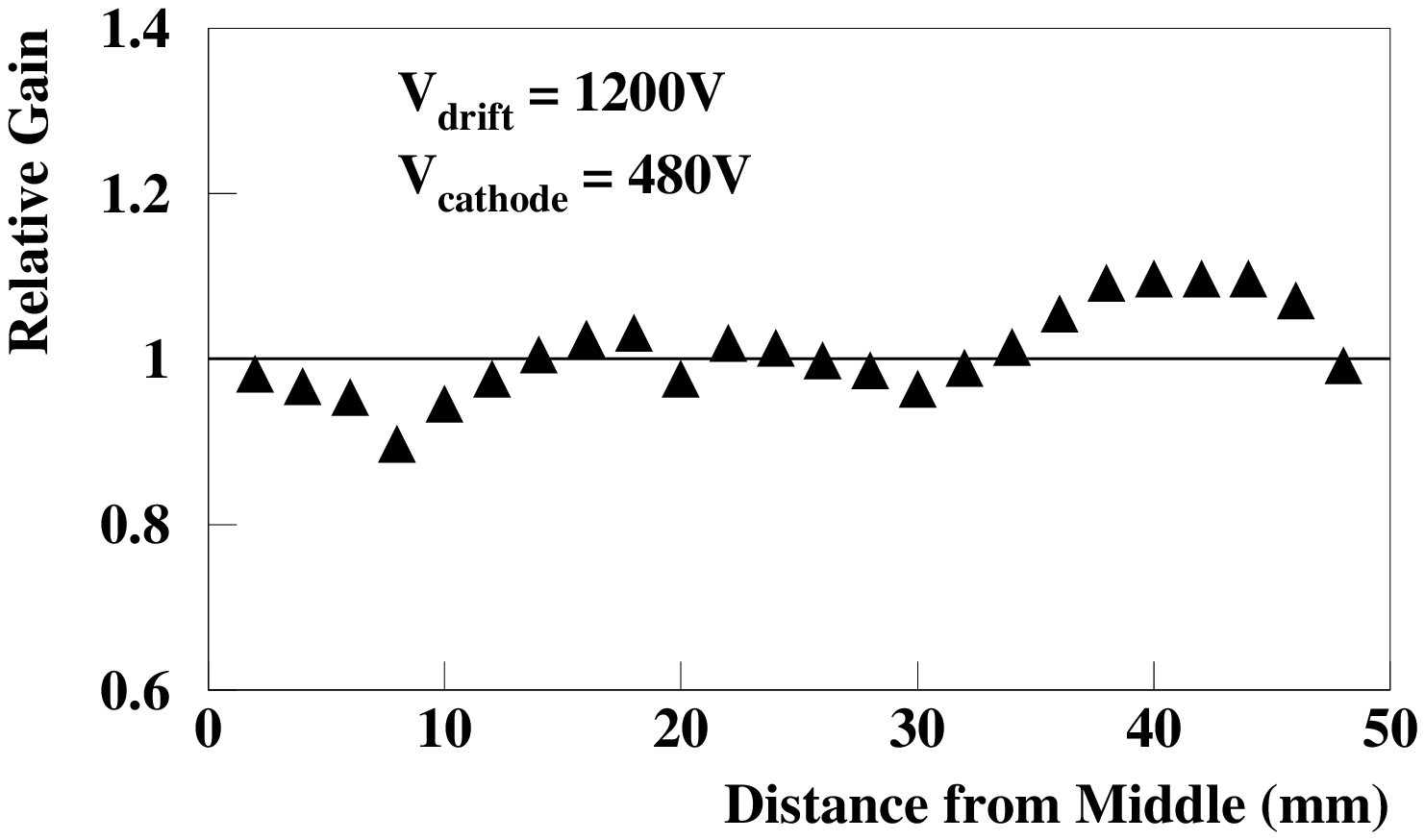,height=8cm}}

Figure 9

\end{center}
\end{figure}

\newpage

\begin{figure}
\begin{center}

\begin{tabular}{m{.5cm}m{9cm}}
a) &
\mbox{\epsfig{file=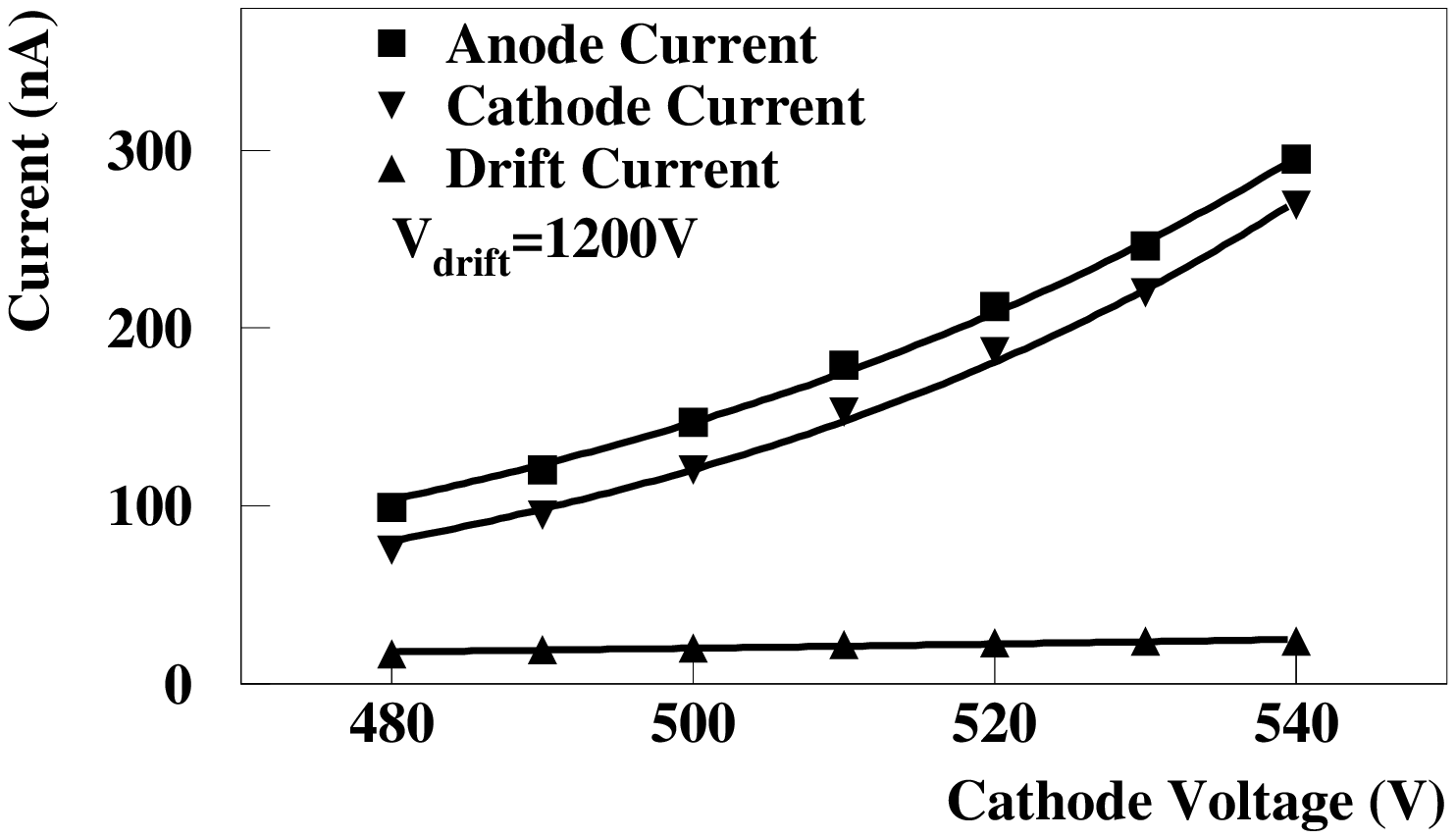,height=8cm}}\\
b) &
\mbox{\epsfig{file=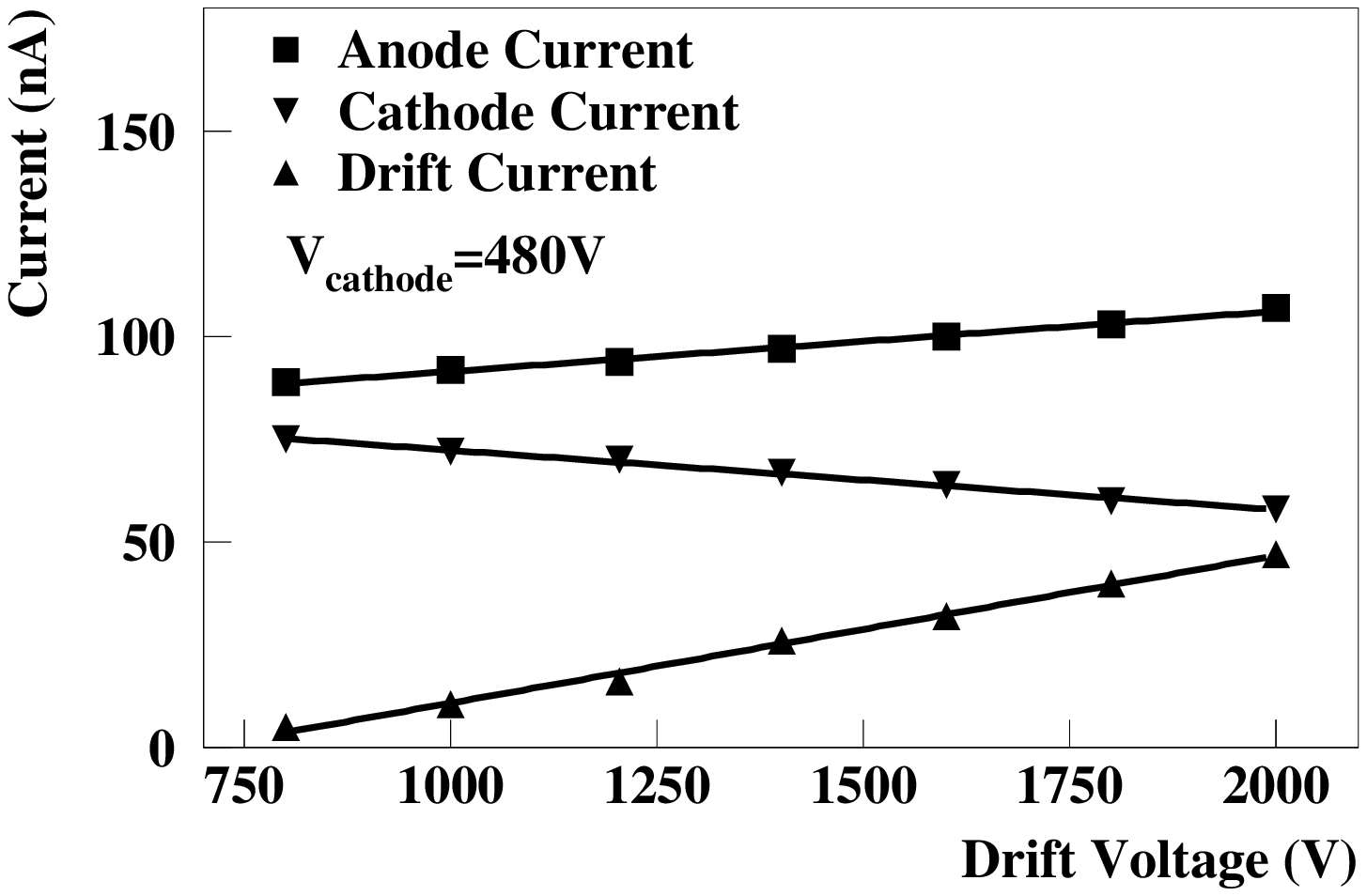,height=8cm}}
\end{tabular}

Figure 10
\end{center}
\end{figure}

\newpage

\begin{figure}
\begin{center}

\begin{turn}{-90}
\mbox{\epsfig{file=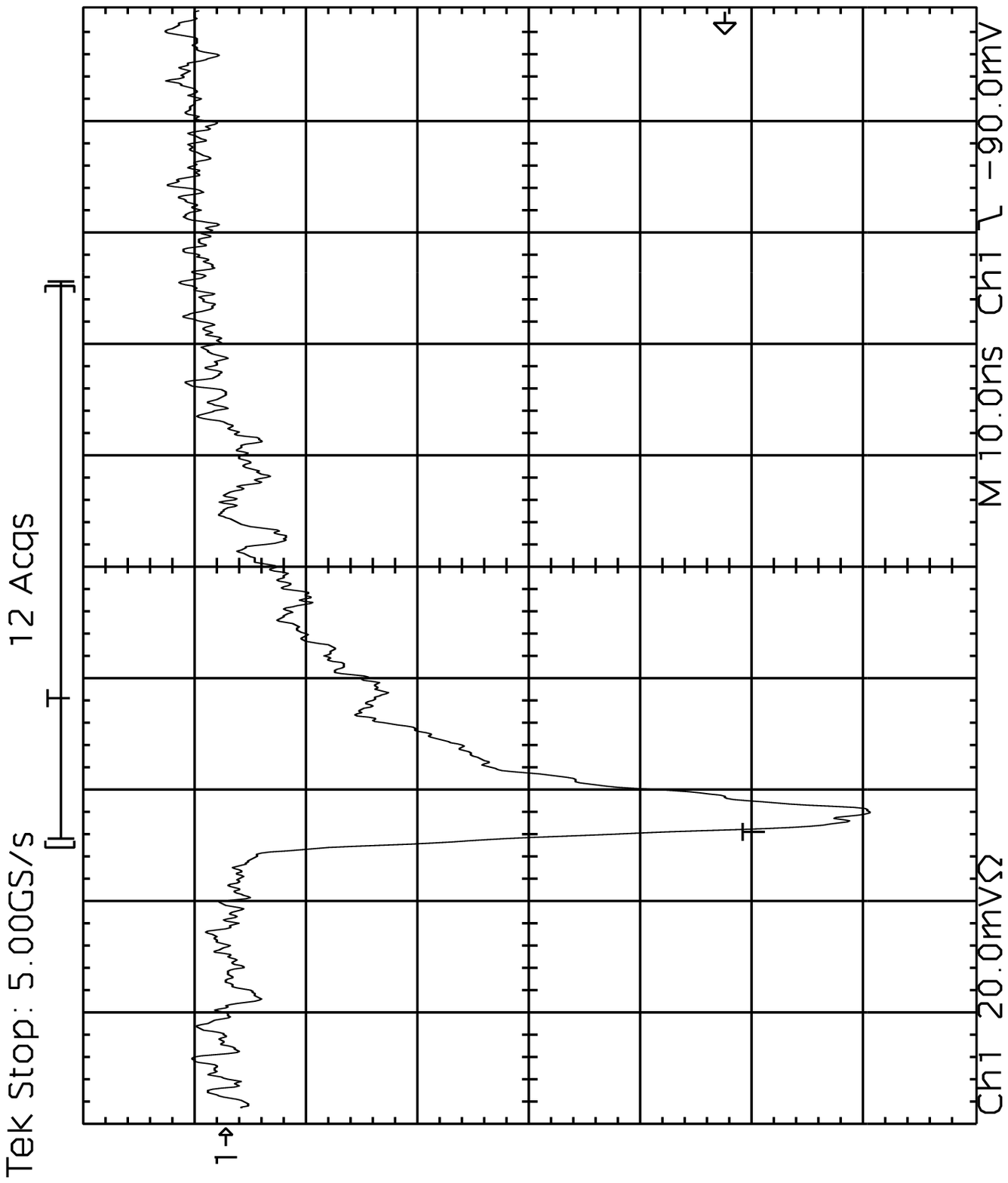,height=10cm}}
\end{turn}

Figure 11

\vspace{15cm}



\end{center}
\end{figure}

\end{document}